\documentclass[10pt,conference]{IEEEtran}
\IEEEoverridecommandlockouts
\usepackage{cite}
\usepackage{optidef}
\usepackage{booktabs}
\usepackage{mathtools}
\usepackage{amsmath}
\usepackage{amsfonts}
\usepackage{amssymb}
\usepackage{amsthm}
\usepackage{caption}
\usepackage{subcaption}
\usepackage{cases}
\usepackage{algorithm,algorithmic}
\usepackage[binary-units=true]{siunitx}
\usepackage{hyperref}
\def\BibTeX{{\rm B\kern-.05em{\sc i\kern-.025em b}\kern-.08em
		T\kern-.1667em\lower.7ex\hbox{E}\kern-.125emX}}

\newcommand\scalemath[2]{\scalebox{#1}{\mbox{\ensuremath{\displaystyle #2}}}}
\newcommand*{\bfrac}[2]{\genfrac{}{}{0pt}{}{#1}{#2}}
\newcommand{\mat}[1]{\ensuremath{{\mathbf{\MakeUppercase{#1}}}}}
\makeatletter
\renewcommand{\vec}[1]{%
	\ifcat\relax\noexpand#1%
	\ensuremath{\boldsymbol{\lowercase{#1}}}%
	\else
	\ensuremath{\mathbf{\lowercase{#1}}}%
	\fi
}
\makeatother
\newcommand{\sumlim}[2]{\ensuremath{\sum\limits_{#1}^{#2}}}

\newcommand{\transpose}[1]{\ensuremath{{#1}^{\textsc{t}}}}
\newcommand{\inverse}[1]{\ensuremath{{#1}^{-1}}}
\newcommand{\R}{\ensuremath{\mathbb{R}}}
\newcommand{\norm}[1]{\left|\left|#1\right|\right|}

\floatname{algorithm}{Algorithm}
\newcommand{\KwIn}[1]{\textbf{Input:} #1}
\newcommand{\KwOut}[1]{\textbf{Output:} #1}
\newcommand{\trace}[1]{\ensuremath{\text{Tr}\left(#1\right)}}

\begin{document}
\title{Stimulus-Informed Generalized Canonical Correlation Analysis of Stimulus-Following Brain Responses
	\thanks{\hspace{-0.27cm}This research is funded by a PDM mandate from KU Leuven (for S. Geirnaert, No PDMT1/22/009), FWO project nr. G081722N, the European Research Council (ERC) under the European Union’s Horizon 2020 research and innovation programme (grant agreement No 802895), and the Flemish Government (AI Research Program). The scientific responsibility is assumed by its authors.}}
\author{\IEEEauthorblockN{Simon Geirnaert\textsuperscript{1,2,3}, Tom Francart\textsuperscript{2,3}, and Alexander Bertrand\textsuperscript{1,3}}\\
	\IEEEauthorblockA{\textsuperscript{1}\textit{KU Leuven, Department of Electrical Engineering (ESAT),}\\ \textit{STADIUS Center for Dynamical Systems, Signal Processing and Data Analytics}, Leuven, Belgium}
	\IEEEauthorblockA{\textsuperscript{2}\textit{KU Leuven, Department of Neurosciences, ExpORL}, Leuven, Belgium}
	\IEEEauthorblockA{\textsuperscript{3}\textit{KU Leuven Institute for Artificial Intelligence (Leuven.AI)}, Leuven, Belgium}
	\IEEEauthorblockA{\{simon.geirnaert,tom.francart,alexander.bertrand\}@kuleuven.be}}

\maketitle

\begin{abstract}
	In brain-computer interface or neuroscience applications, generalized canonical correlation analysis (GCCA) is often used to extract correlated signal components in the neural activity of different subjects attending to the same stimulus. This allows quantifying the so-called inter-subject correlation or boosting the signal-to-noise ratio of the stimulus-following brain responses with respect to other (non-)neural activity. GCCA is, however, stimulus-unaware: it does not take the stimulus information into account and does therefore not cope well with lower amounts of data or smaller groups of subjects. We propose a novel stimulus-informed GCCA algorithm based on the MAXVAR-GCCA framework. We show the superiority of the proposed stimulus-informed GCCA method based on the inter-subject correlation between electroencephalography responses of a group of subjects listening to the same speech stimulus, especially for lower amounts of data or smaller groups of subjects.
\end{abstract}


\section{Introduction}
\label{sec:intro}

Traditionally, several neuroscience applications and brain-computer interface technologies heavily rely on active participation of the user following specific instructions. Moreover, synthetic and controlled stimuli such as flickering visual patterns or beep tones are used, as they evoke more controlled brain responses. Lastly, they rely on multi-trial experiments, where the same controlled stimulus is repeatedly presented to allow enhancing the signal-to-noise ratio (SNR) by averaging the responses across multiple trials~\cite{nicolasAlonso2012brain}. In the past few years, a shift has been occurring towards passive, single-trial experiments using natural sensory stimuli, such as speech or video footage~\cite{geirnaert2021eegBased,geirnaert2022timeAdaptive,poulsen2017eeg,dmochowski2012correlated,dmochowski2014audience,ki2016attention,vanthornhout2019effect}. This shift opens doors to novel applications, for example, in attention tracking in hearing devices~\cite{geirnaert2021eegBased,geirnaert2022timeAdaptive}, the (online) classroom~\cite{poulsen2017eeg}, neuromarketing~\cite{dmochowski2012correlated,dmochowski2014audience}, or virtual reality environments~\cite{delvigne2021attention}. Given that we are interested in stimulus-following brain responses, we focus on electroencephalography (EEG) or magnetoencephalography (MEG), which have an excellent temporal resolution~\cite{nicolasAlonso2012brain}.

This shift brings along several signal processing-related challenges, such as the strong subject-specificity arising from using uncontrolled and natural stimuli that evoke highly variable responses, and especially the low SNR of the stimulus-following EEG responses buried under interfering non-neural and neural activity that is not time-locked to the stimulus. In this single-trial setting, the low SNR can not be dealt with anymore by averaging multiple trials, such that data-driven filtering methods are required to enhance the stimulus-following neural response and suppress all other noise sources.

In this paper, we focus on generalized canonical correlation analysis (GCCA), which allows extracting correlated components across multiple (EEG) recordings, for example from different subjects. GCCA can then be used not only to enhance the SNR but also to quantify inter-subject correlation (ISC)~\cite{dmochowski2012correlated,dmochowski2014audience,ki2016attention,poulsen2017eeg}, for dimensionality reduction, or to summarize a set of EEG recordings~\cite{deCheveigne2019multiway}. An overview of GCCA for brain data analysis can be found in de Cheveigné et al.~\cite{deCheveigne2019multiway}. 

We focus on the problem of extracting the stimulus-following neural response from a set of EEG recordings from multiple subjects attending to the same stimulus. One of the strengths of GCCA here is that it is \emph{stimulus-unaware}: it makes no assumptions about the stimulus (representation) and also works when the stimulus is unavailable. However, this strength can be a weakness at the same time. As explained before, one of the main challenges is the notoriously low SNR of the stimulus-following neural response. Maximally exploiting all available side information, including the stimulus, is paramount to optimally boost the SNR, even more so when only a few subjects or small amount of data are available. The latter regularly occurs, for example, in a time-adaptive context for online processing~\cite{geirnaert2022timeAdaptive}. Therefore, we modify the GCCA formulation to include the stimulus as side-information, leading to \emph{stimulus-informed} GCCA (SI-GCCA).

In Section~\ref{sec:maxvar-gcca}, we revisit GCCA and, more specifically, its MAXVAR formulation. We then present the novel SI-GCCA method in Section~\ref{sec:si-gcca}. In Section~\ref{sec:experiments}, we compare GCCA and SI-GCCA. Conclusions are drawn in Section~\ref{sec:conclusion}.

\section{MAXVAR-GCCA}
\label{sec:maxvar-gcca}

Several ways exist to generalize traditional CCA to multiple views or subjects, of which SUMCORR and MAXVAR are the most popular formulations~\cite{kettenring1971canonical}. Here we choose the MAXVAR-GCCA formulation because it allows for an easy introduction of the stimulus information (Section~\ref{sec:si-gcca}) and has the nice property of leading to a generalized eigenvalue decomposition (GEVD)~\cite{kettenring1971canonical}.

Consider $K$ stimulus-driven zero-mean EEG signals $\mat{X}_k \in \R^{T \times M}$ of $K$ subjects attending to the same stimulus, for example, a speech signal. $T$ denotes the number of EEG samples, where each sample is $M$-dimensional\footnote{In principle, a different dimensionality per subject can be used.}, arising from, for example, different EEG channels and/or time-lagged copies of each channel. In MAXVAR-GCCA, the goal is to find the $M \times Q$-dimensional filters $\mat{W}_k \in \R^{M \times Q}$ and a $Q$-dimensional shared signal subspace spanned by $\mat{S} = \begin{bmatrix} \vec{s}_1 & \cdots & \vec{s}_Q \end{bmatrix} \in \R^{T\times Q}$ such that the latter is on average closest to the transformed individual EEG signals $\mat{X}_k\mat{W}_k$~\cite{kettenring1971canonical,via2007learning,hovine2021distributed}:
\begin{mini}|s|
	{\substack{\mat{W}_1, \dots, \mat{W}_K,\mat{S}}}{\sumlim{k=1}{K}\norm{\mat{S}-\mat{X}_k\mat{W}_k}_F^2}{\label{eq:maxvar-opt}}{}
	\addConstraint{\transpose{\mat{S}}\mat{S} = \mat{I}_Q,}
\end{mini}
with $\mat{I}_Q$ the $Q$-dimensional identity matrix.

The Lagrangian function is then equal to:
\[
\begin{split}
	\mathcal{L}&\!\left(\mat{W}_1,\dots,\mat{W}_K,\mat{S},\mat{\Lambda}\right) =\\
	&\quad  K\trace{\transpose{\mat{S}}\mat{S}}-2\sumlim{k=1}{K}\trace{\transpose{\mat{S}}\mat{X}_k\mat{W}_k}+ \\
	&\quad  \sumlim{k=1}{K}\trace{\transpose{\mat{W}}_k\transpose{\mat{X}}_k\mat{X}_k\mat{W}_k} -\trace{\left(\transpose{\mat{S}}\mat{S}-\mat{I}_Q\right)\mat{\Lambda}}, \\
\end{split}
\]
with $\mat{\Lambda} \in \R^{Q \times Q}$ a symmetric matrix containing the Lagrange multipliers. From the Karush-Kuhn-Tucker conditions, we find the following three equations:
\begin{numcases}{}
	\transpose{\mat{X}}_k\mat{S} = \transpose{\mat{X}}_k\mat{X}_k\mat{W}_k, \; k = 1,\dots,K\label{eq:kkt-1}\\
	\mat{s} = \sumlim{k=1}{K}\mat{X}_k\mat{W}_k\mat{\Omega},\label{eq:s}\\
	\transpose{\mat{s}}\mat{s} = \mat{I}_Q\label{eq:kkt-3},
\end{numcases}       
with $\mat{\Omega} = \inverse{\left(K\mat{I}_Q-\mat{\Lambda}\right)}$ also a symmetric matrix. Defining $\mat{R}_{kl} = \transpose{\mat{X}}_k\mat{X}_l \in \R^{M \times M}$ as the sample crosscorrelation matrix of $\mat{X}_k$ and $\mat{X}_l$, and plugging \eqref{eq:s} into \eqref{eq:kkt-1}, one obtains 
\begin{equation}
	\label{eq:almost-gevd}
	\mat{R}_{D_{xx}}\mat{W} = \mat{R}_{xx}\mat{w}\mat{\Omega},
\end{equation}
with $\mat{W} = \transpose{\begin{bmatrix}  \transpose{\mat{w}}_1 & \cdots & \transpose{\mat{w}}_K\end{bmatrix}} \in \R^{KM \times Q}$, $\mat{R}_{D_{xx}} = \linebreak\text{Blkdiag}\!\left(\mat{R}_{11},\dots,\mat{R}_{KK}\right) \in \R^{KM \times KM}$ the block diagonal matrix containing the per-subject autocorrelation matrices, and $\mat{R}_{xx} = \transpose{\mat{X}}\mat{X} \in \R^{KM \times KM}$, with $\mat{X} = \begin{bmatrix} \mat{X}_1 & \cdots & \mat{X}_K \end{bmatrix} \in \R^{T \times KM}$.

Given that $\mat{\Omega}$ is a symmetric matrix, it is orthogonally diagonalizable as
\begin{equation}
	\label{eq:omega-orth}
	\mat{\Omega} = \mat{U}\mat{\Sigma}\transpose{\mat{U}},
\end{equation}
with $\mat{U} \in \R^{Q \times Q}$ an orthogonal matrix and $\mat{\Sigma} \in \R^{Q\times Q}$ a diagonal matrix. Substituting \eqref{eq:omega-orth} in \eqref{eq:almost-gevd} leads to the GEVD that gives the solution:
\[
\mat{R}_{D_{xx}}\mat{W}\mat{U} = \mat{R}_{xx}\mat{W}\mat{U}\mat{\Sigma},
\]
The optimal filters in $\mat{W}$ are then in the subspace spanned by the $Q$ generalized eigenvectors (GEVcs) corresponding to the smallest generalized eigenvalues (GEVls)\footnote{It can be found that the objective function at the optimal solution is equal to $K-\trace{\inverse{\mat{\Sigma}}}$.}. As it can be shown that the solution is defined upon any orthogonal transformation~\cite{ghojogh2022eigenvalue}\footnote{Proof omitted due to space constraints.}, for convenience, we take the GEVcs themselves as the solution. Note that this corresponds to selecting the matrix of Lagrange multipliers $\mat{\Lambda}$, or $\mat{\Omega}$, to be diagonal. The correct scaling of the GEVcs can be retrieved from \eqref{eq:kkt-3}. Lastly, as shown in~\cite{via2007learning,hovine2021distributed}, using \eqref{eq:s}, the problem in~\eqref{eq:maxvar-opt} is equivalent to minimizing the sum of the pairwise distances between $\mat{X}_k\vec{w}_k$ and $\mat{X}_l\vec{w}_l$ for all $k$ and $l$, relating to the intuitive definition of GCCA, i.e., maximizing pairwise correlations.

\section{SI-GCCA}
\label{sec:si-gcca}

In stimulus-informed GCCA (SI-GCCA), we now want to use the stimulus, if available, as side information to steer the estimation of the shared signal subspace and filters. Such a framework is based on the (reasonable) assumption that the correlated components across the $K$ EEG recordings correspond to the neural responses to the common stimulus. Using the stimulus to inform the GCCA estimation can be viewed as a task-informed regularization technique to cope with situations where only little data is available (such as in an adaptive or online context) or only a few subjects are available.

Assume a $P$-dimensional stimulus representation $\mat{Y} \in \R^{T \times P}$ (see Section~\ref{sec:filter-design} for an example of such a representation for a speech stimulus). We propose to introduce the stimulus information into the MAXVAR-GCCA optimization problem in~\eqref{eq:maxvar-opt} by adding a forward model/encoder $\mat{V} \in \R^{P \times Q}$ on the stimulus to the shared signal subspace as an extra term:
\begin{mini}|s|
	{\substack{\mat{W}_1, \dots, \mat{W}_K,\mat{V},\mat{S}}}{\sumlim{k=1}{K}\norm{\mat{S}-\mat{X}_k\mat{W}_k}_F^2+\rho\norm{\mat{s}-\mat{Y}\mat{v}}_F^2}{\label{eq:sigcca-opt}}{}
	\addConstraint{\transpose{\mat{S}}\mat{S} = \mat{I}_Q.}
\end{mini}
This formulation has the advantage that it remains a MAXVAR-GCCA problem with the stimulus as an additional view, so it can still be solved as a GEVD. The additional term related to the stimulus `pulls' on the shared signal subspace, where the hyperparameter $\rho$ determines the weight.

The new MAXVAR-GCCA optimization problem in~\eqref{eq:sigcca-opt} can be solved similarly to Section~\ref{sec:maxvar-gcca} by defining the new variables $\tilde{\mat{w}} = \transpose{\begin{bmatrix}  \transpose{\mat{w}}_1 & \cdots & \transpose{\mat{w}}_K & \transpose{\mat{v}} \end{bmatrix}} \in \R^{(KM+P) \times Q}$ and $\tilde{\mat{X}} =\linebreak \begin{bmatrix} \mat{X}_1 & \cdots & \mat{X}_K & \mat{Y} \end{bmatrix} \in \R^{T \times (KM+P)}$. We now a priori select the matrix of Lagrange multipliers $\mat{\Lambda}$ to be diagonal. The resulting shared signal subspace $\mat{S}$ and corresponding GEVD then become:
\begin{numcases}{}
	\mat{S} = \left(\sumlim{k=1}{K}\mat{X}_k\mat{W}_k+\rho\mat{Y}\mat{V}\right)\mat{\Omega}\label{eq:s-sigcca}\\
	\mat{R}_{D_{\tilde{x}\tilde{x}}}\tilde{\mat{W}} = \tilde{\mat{R}}_{\tilde{x}\tilde{x}}\tilde{\mat{W}}\mat{\Omega},
\end{numcases}
with $\mat{R}_{D_{\tilde{x}\tilde{x}}} = \text{Blkdiag}\!\left(\mat{R}_{11},\dots,\mat{R}_{KK},\mat{R}_{yy}\right), \mat{\Omega} = \linebreak\inverse{\left((K+\rho)\mat{I}_Q-\mat{\Lambda}\right)}$ again a diagonal matrix, and 
\begin{equation}
	\label{eq:R-mat}
	\tilde{\mat{R}}_{\tilde{x}\tilde{x}} = \begin{bmatrix} 
		\mat{R}_{11} & \dots & \mat{R}_{1K} & \rho\mat{R}_{1y} \\
		\vdots & & \vdots & \vdots \\
		\mat{R}_{1K} & \dots & \mat{R}_{KK} & \rho\mat{R}_{Ky} \\
		\mat{R}_{1y} & \dots & \mat{R}_{Ky} & \rho\mat{R}_{yy} \\
	\end{bmatrix}.\end{equation}

By taking the GEVcs corresponding to the smallest GEVls, the correct filter subspace can again be retrieved. The correct scaling of the GEVcs can, as before, be retrieved from the equality constraint $\transpose{\mat{S}}\mat{S} = \mat{I}_Q$. The SI-GCCA algorithm is summarized in Algorithm~\ref{algo:si-gcca}, while a MATLAB implementation is available at \url{https://github.com/AlexanderBertrandLab/si-gcca}.

\begin{algorithm}
	\caption{SI-GCCA}
	\label{algo:si-gcca}
	\KwIn{$K$ stimulus-driven EEG signals $\mat{X}_k \in \R^{T \times M}$, stimulus $\mat{Y}\in \R^{T \times P}$, hyperparameter $\rho$, subspace dimension $Q$}\\
	\KwOut{\makebox[0.1\linewidth][l]{Per-subject filters $\mat{W}_k\!\in \R^{M \times Q}$}}
	\begin{algorithmic}[1]
		\STATE Compute correlation matrices $\tilde{\mat{R}}_{\tilde{x}\tilde{x}}$ as in~\eqref{eq:R-mat} and $\mat{R}_{D_{\tilde{x}\tilde{x}}} = \text{Blkdiag}\!\left(\mat{R}_{11},\dots,\mat{R}_{KK},\mat{R}_{yy}\right),$	with $\mat{R}_{kl} = \transpose{\mat{X}}_k\mat{X}_l$ (optionally with regularization)
		\STATE Compute $\tilde{\mat{W}}$ as the $Q$ GEVcs corresponding to the $Q$ smallest GEVls of the matrix pencil $\left(\mat{R}_{D_{\tilde{x}\tilde{x}}},\tilde{\mat{R}}_{\tilde{x}\tilde{x}}\right)$
		\STATE Scale the GEVcs such that $\transpose{\mat{S}}\mat{S} = \mat{I}_Q$, with $\mat{S}$ defined in \eqref{eq:s-sigcca}
		\STATE Extract $\mat{W}_k$ from $\tilde{\mat{W}} = \transpose{\begin{bmatrix}  \transpose{\mat{w}}_1 & \cdots & \transpose{\mat{w}}_K & \transpose{\mat{v}} \end{bmatrix}}$
	\end{algorithmic}
\end{algorithm}

\section{Experiments and results}
\label{sec:experiments}

\subsection{Task}
\label{sec:task}

We evaluate and compare the GCCA and SI-GCCA methods in the context of a group of subjects listening to the same speech stimulus, i.e., a story. For simplicity, we compute only one filter per subject, i.e., $Q = 1$. In this case, the set of $Q$ filters $\mat{W}_k$ per subject boil down to a single $M$-dimensional filter $\vec{w}_k \in \R^{M}$ per subject. The goal is to use (SI-)GCCA to quantify the inter-subject correlation (ISC), i.e., the correlation between the (transformed) EEG recordings of the different subjects, which could serve as a proxy for attentional engagement of the group~\cite{dmochowski2012correlated,dmochowski2014audience,ki2016attention,poulsen2017eeg}. Given the zero-mean (one-dimensional) (SI-)GCCA transformed EEG signals $\vec{z}_k = \mat{X}_k\vec{w}_k \in \R^{T}$, this ISC is defined as the average pairwise Pearson correlation coefficient:
\begin{equation}
	\label{eq:isc}
	\text{ISC} = \frac{2}{K(K-1)}\sumlim{k=1}{K-1}\sumlim{l=k+1}{K}\frac{\transpose{\vec{z}}_k\vec{z}_l}{\norm{\vec{z}_k}_2\norm{\vec{z}_l}_2}.
\end{equation}
This ISC, evaluated on trials of $\SI{60}{\second}$ on an independent test set, serves as the performance metric. The higher the ISC the better, as this means that the neural processes unrelated to the stimulus are more suppressed.

In Section~\ref{sec:dataset}, we briefly introduce the dataset, while the design choices are shown in Section~\ref{sec:design-choices}. The GCCA and SI-GCCA method are compared in two different experiments: varying the amount of training data (Section~\ref{sec:atd}) and the number of subjects (Section~\ref{sec:as}). MATLAB code for these experiments is available at \url{https://github.com/AlexanderBertrandLab/si-gcca}.

\subsection{Dataset}
\label{sec:dataset}
The dataset is taken from Broderick et al.~\cite{broderick2018electrophysiological} and is publicly available~\cite{broderick_michael_p_2019_5080270}. We use the EEG data of the first experiment, containing data from 19 normal-hearing subjects, all listening to the same narrative speech, i.e., an audiobook. This audiobook was presented in $20$ trials, each of length around $\SI{3}{\minute}$. Each of these trials was cut into new trials of $\SI{60}{\second}$, resulting in $\SI{52}{\minute}$ of synchronized EEG/speech data per subject. The EEG is recorded using a $128$-channel BioSemi ActiveTwo system and re-referenced to the average mastoid channel.

\subsection{Design choices}
\label{sec:design-choices}

\subsubsection{Stimulus representation and bandpass filtering}

Given that it has been repeatedly established that EEG signals track the low-frequency envelope of the attended speech signal~\cite{geirnaert2021eegBased,geirnaert2022timeAdaptive,diLiberto2015low}, the speech stimulus in the SI-GCCA framework is represented by the amplitude envelope, computed using the Hilbert transform~\cite{diLiberto2015low}. Furthermore, both the speech envelope and EEG signals are filtered in the $\delta$-band, i.e., between $\SIrange{1}{4}{\hertz}$, which has shown to provide good results in the context of tracking auditory attention to a speech signal~\cite{vanthornhout2019effect}. The EEG signals and speech envelope are correspondingly downsampled to $\SI{8}{\hertz}$.

\subsubsection{Filter design}
\label{sec:filter-design}
We will estimate spatio-temporal filters on the $C = 128$-channel EEG signals of the different subjects, augmented with $L$ time lags to compensate for temporal differences in processing between subjects. Therefore, $M=CL$, with $L = 5$, i.e., the time lags are ranging from $-\frac{L-1}{2} = -2$ to $\frac{L-1}{2} = 2$, corresponding to an integration window of $[-250,250]\SI{}{\milli\second}$. The resulting EEG regression matrix $\mat{X}_k$ then becomes a block Hankel matrix:
\[
\mat{X}_k = \begin{bmatrix}
	\mat{X}_{k,1} & \dots & \mat{X}_{k,128}
\end{bmatrix},\]
\[
\mat{X}_{k,c} = \scalemath{0.81}{\begin{bmatrix}
		0 & 0 & x_{k,c}(0) & x_{k,c}(1) & x_{k,c}(2) \\
		0 & x_{k,c}(0) & x_{k,c}(1) & x_{k,c}(2) & x_{k,c}(3) \\
		x_{k,c}(0) & x_{k,c}(1) & x_{k,c}(2) & x_{k,c}(3) & x_{k,c}(4)\\
		\vdots & \vdots & \vdots & \vdots & \vdots \\
		x_{k,c}(T-3) & x_{k,c}(T-2) & x_{k,c}(T-1) & 0 & 0\\
	\end{bmatrix},}
\]
with $x_{k,c}(t)$ the $c$\textsuperscript{th}-channel EEG signal of the $k$\textsuperscript{th} subject. 

In the SI-GCCA estimation, the stimulus representation is the speech envelope and $10$ time-lagged copies of it to allow for spectral filtering (including compensation for the intrinsic delay between the stimulus and EEG response) by $\vec{v}$, leading to a pre-stimulus integration window of $[-1.25,0]\SI{}{\second}$~\cite{geirnaert2021eegBased}. Therefore, the stimulus representation $\mat{Y}$ is also a Hankel matrix with $P = 11$ columns.

\subsubsection{Correlation matrix estimation}
All $\SI{60}{\second}$-trials are a priori normalized by setting the mean per channel to zero and putting the Frobenius norm across all channels to one. Furthermore, all correlation matrices ($\mat{R}_{xx},\tilde{\mat{R}}_{\tilde{x}\tilde{x}},\dots$) are estimated using ridge regression, where the regularization parameter is automatically estimated using the method proposed by Ledoit and Wolf~\cite{ledoit2004well}.

\subsubsection{Hyperparameter selection}
The hyperparameter $\rho$ in the SI-GCCA estimation problem~\eqref{eq:sigcca-opt}, which determines the weight put on the stimulus, is selected based on the ISC using an independent validation set consisting of $20\%$ of the trials in the test set (which are thereafter removed from this test set and not used in the ISC computation). A sweep of hyperparameters in the range of $\{0,10^{-1},10^{-0.5},\dots,10^{3}\}$ is performed and the $\rho$ leading to the highest ISC on the validation set is selected.

\subsubsection{Significance level computation and statistical comparison}
\label{sec:stat}
In Section~\ref{sec:atd} and \ref{sec:as}, per ISC, the significance level is determined via a random permutation test: for the GCCA algorithm, a random 5-fold cross-validation is performed. The resulting GCCA output signals are then $\num{10 000}$ times randomly permuted per $\SI{60}{\second}$-trial across the different subjects (i.e., removing correlated patterns), after which the $5\%$ significance level can be computed. This significance level is mainly determined by the trial length and number of subjects used to compute the ISC.

The statistical comparison between the (SI-)GCCA methods is performed using a linear mixed-effect model (LMEM), where the method is modelled as a fixed effect and the amount of training data (Section~\ref{sec:atd}) or number of subjects (Section~\ref{sec:as}) as random slope (i.e., the ISC can vary differently per method across the random effect):
\[
\text{ISC} \sim 1 + \text{method} + \left(\text{method}\left\rvert\scalemath{0.81}{\bfrac{\text{amount of training data}}{\text{number of subjects}}}\right.\right).
\]

\subsection{Amount of training data}
\label{sec:atd}
In this experiment, we compare the GCCA and SI-GCCA methods for various amounts of training data while using all 19 subjects to train and test. The amount of training data is varied between $1$ and $\SI{45}{\minute}$. Per amount of training data, $50$ Monte Carlo runs are performed, in which the $\SI{60}{\second}$-training trials are randomly sampled from the $\SI{52}{\minute}$ available data. The remaining $\SI{60}{\second}$-trials are used to test the resulting decoders. In case of the SI-GCCA method, this rest is split into $80\%$ test data and $20\%$ validation data.

\begin{figure}
	\centering	
	\includegraphics[width=1\linewidth]{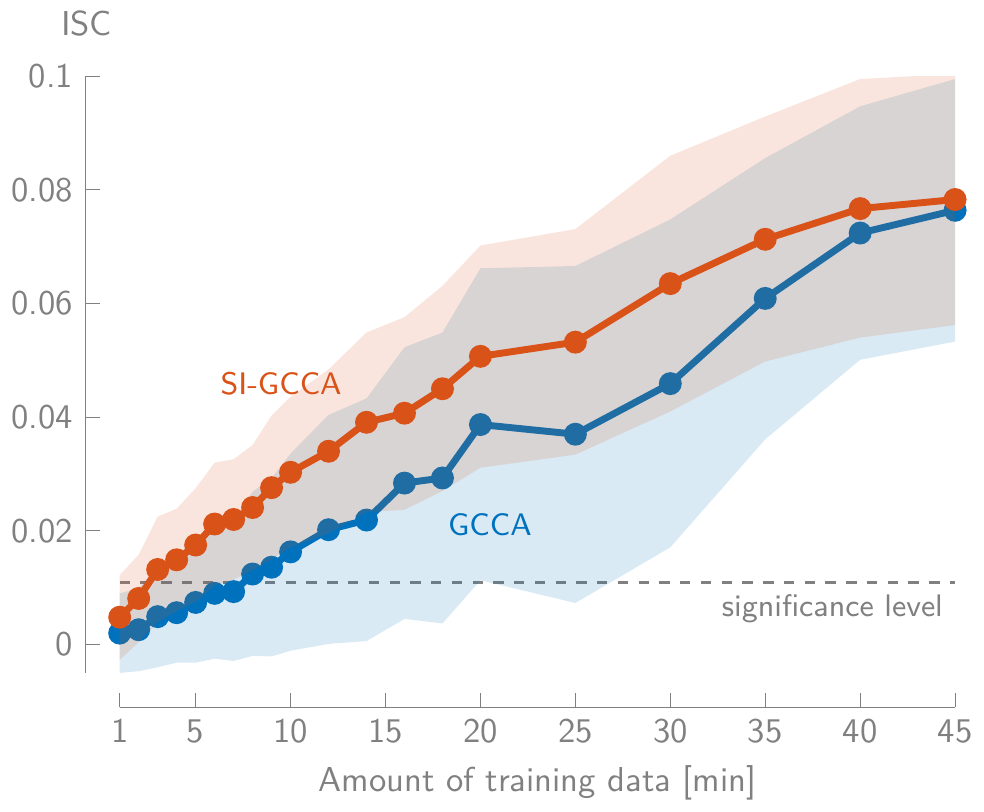}
	\caption{SI-GCCA outperforms GCCA across all amounts of training data (mean $\pm$ standard deviation across all test trials and Monte Carlo runs).}
	\label{fig:atd}
\end{figure}

Figure~\ref{fig:atd} shows the ISC as a function of the amount of training data used. The SI-GCCA method outperforms the GCCA method across all amounts of training data (mean difference in ISC of $0.011$), which is confirmed using the LMEM (significant difference, $p$-value $<0.001$). This shows that the stimulus information can effectively improve the per-subject filter estimation. The difference becomes smaller only when large amounts of training data are used: the side-information introduced by the stimulus regularizer decreases as the high dimensionality of the problem can be compensated by the large amounts of training data. 

Lastly, the SI-GCCA method allows retrieving significant ISCs already with $\SI{3}{\minute}$ of training data (as opposed to $\SI{8}{\minute}$ for the GCCA method), which is a crucial advantage in online, adaptive BCI applications.

\subsection{Number of subjects}
\label{sec:as}
We also compare the GCCA and SI-GCCA methods for varying numbers of subjects while using $5$-fold cross-validation on all available data (corresponding to $\SI{42}{\minute}$ of training data). Per number of subjects, $25$ different combinations are sampled from all 19 subjects. 

\begin{figure}
	\centering	
	\includegraphics[width=1\linewidth]{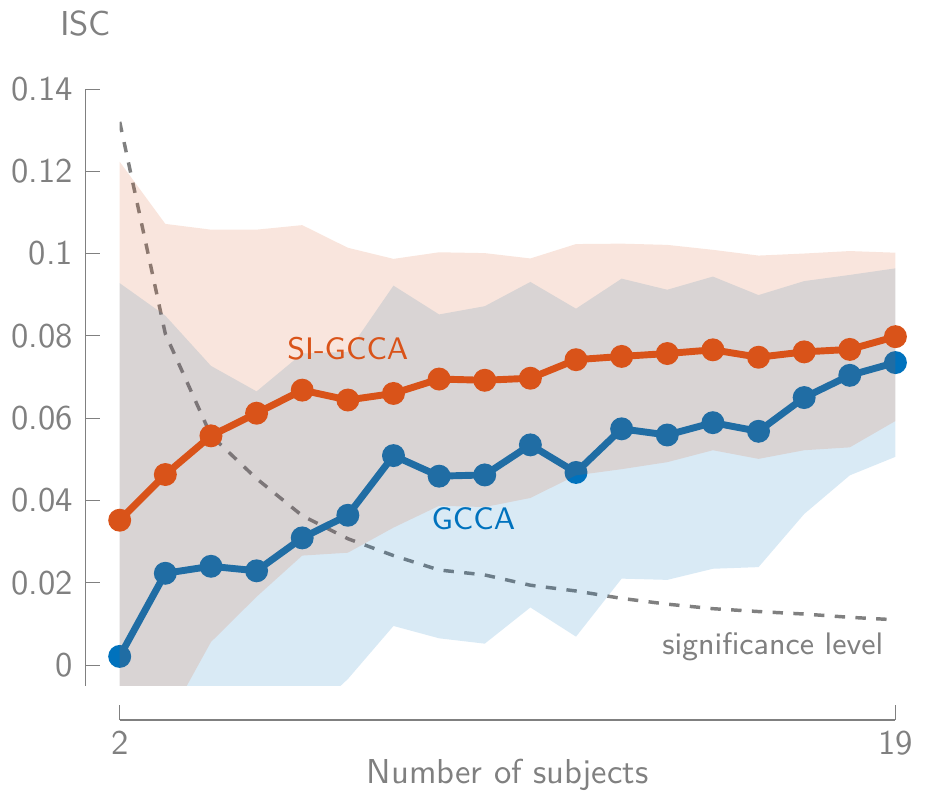}
	\caption{SI-GCCA outperforms GCCA across all numbers of subjects (mean $\pm$ standard deviation across all folds and different sampled combinations).}
	\label{fig:as}
\end{figure}

Figure~\ref{fig:as} shows the ISC as a function of the number of subjects. The more subjects are used, the lower the significance level, as averaging more pairwise correlations in the ISC counteracts spurious correlations. The SI-GCCA method effectively compensates for the lower number of subjects, where less information is available to extract the shared stimulus-related activity. Using an LMEM, there is a significant difference between both methods across all numbers of subjects (mean difference in ISC of $0.023$, $p$-value $<0.001$). Furthermore, SI-GCCA results in significant correlations when more than $5$ subjects are present, which is only the case for $7$ subjects for the GCCA method.

\section{Conclusion}
\label{sec:conclusion}

In this paper, we have proposed a novel stimulus-informed GCCA framework that retains the attractive properties of the GCCA-MAXVAR framework, i.e., it can still be computed as a GEVD. We have demonstrated its superiority w.r.t. GCCA when quantifying the ISC between subjects listening to the speech stimulus, especially when smaller amounts of training data and fewer subjects are available.

\section{Acknowledgements}

The authors would like to thank Thomas Strypsteen and Miguel Bhagubai for their help in providing the necessary computing power and Charles Hovine for the brainstorming sessions and proofs about the invariance of the solution upon an orthogonal transformation.

\bibliographystyle{ieeetran}
\bibliography{papers-ref-si-gcca}
\end{document}